\documentclass[useAMS,usenatbib]{mn2e}
\usepackage{graphicx}
\usepackage{rotating}
\usepackage{longtable}
\usepackage{lscape}
\usepackage[usenames]{color}

\def\Hb{H$\beta$}
\def\lsim{\mathrel{\rlap{\lower 3pt \hbox{$\sim$}} \raise 2.0pt \hbox{$<$}}}
\def\gsim{\mathrel{\rlap{\lower 3pt \hbox{$\sim$}} \raise 2.0pt \hbox{$>$}}}

\title[A black hole binary in SDSSJ0927]{SDSSJ092712.65+294344.0: a candidate massive black hole binary}

\author[Dotti et al.]{M. Dotti,$^{1}$\thanks{mdotti@umich.edu} C. Montuori,$^{2}$ R. Decarli,$^{3}$ M. Volonteri,$^{1}$ M. Colpi,$^{2}$ F. Haardt$^{3}$\\
             $^1$ Department of Astronomy, University of Michigan, 
             Ann Arbor, MI 48109, USA\\
             $^2$ Universit\`a degli Studi di Milano-Bicocca, Piazza della
             Scienza 3, 20126 Milano, Italy\\
             $^3$ Universit\`a degli Studi dell'Insubria, Via Valleggio 11,
             22100 Como, Italy }

\pagerange{\pageref{firstpage}--\pageref{lastpage}} \pubyear{2008}

\begin{document}	
\maketitle
\label{firstpage}

\begin{abstract}
In this Letter we explore the hypothesis that the quasar
SDSSJ092712.65+294344.0 is hosting a massive black hole binary
embedded in a circumbinary disc. The lightest, secondary black hole is
active, and gas orbiting around it is responsible for the
blue--shifted broad emission lines with velocity off-set of 2650 km
s$^{-1}$, relative to the galaxy rest frame. As the tidal interaction
of the binary with the outer disc is expected to excavate a gap, the
blue-shifted narrow emission lines are consistent with being emitted
from the low-density inhomogeneous gas of the hollow region. From the
observations we infer a binary mass ratio $q\approx 0.3$, a mass for
the primary of $M_1\approx 2 \times 10^9\,M_{\odot},$ and a semi-major
axis of 0.34 pc, corresponding to an orbital period of 370 years. We
use the results of cosmological merger trees to estimate the
likely-hood of observing SDSSJ092712.65+294344.0 as recoiling black
hole or as a binary.  We find that the binary hypothesis is preferred
being one hundred times more probable than the ejection hypothesis. If
SDSSJ092712.65+294344.0 hosts a binary, it would be the one closest
massive black hole binary system ever discovered.
\end{abstract}

\section{Introduction}

If massive black holes (MBHs) are ubiquitous in galaxy spheroids and
galaxies experience multiple mergers during their cosmic assembly, MBH
{\it binaries} (MBHBs hereon) should be common, albeit transient
features of galactic bulges.  Observationally, the paucity of dual
active nuclei points toward rapid inspiral of the MBHs down to parsec
scales where they form a gravitationally bound pair (Mayer et
al. 2007).  After birth, the MBHB further hardens under the action of
gas/dynamical and/or stellar torques (Begelman, Blandford \& Rees
1980) and depending on their efficiency (Milosavljevic \& Merritt
2001; Armitage \& Natarajan 2002; Berczik et al. 2006; Sesana et
al. 2007; Perets et al. 2007; Colpi et al. 2007) the binary may remain
in its hard state for a relatively long time, before coalescing under
the action of gravitational waves.

MBHBs can coalesce shortly, and a clear prediction of general
relativity is that the relic MBH receives a kick in response to the
asymmetric pattern of gravitational waves emitted just prior
coalescence.  For slowly spinning MBHs recoil velocities of a few
$\times 100$ km s$^{-1}$ were predicted, but recent numerical advances
in general relativity have indicated that velocities as extreme as
$\sim 4000$ km s$^{-1}$ can be attained for maximally rotating Kerr
MBHs with spin vectors lying in the orbital plane (Schnittman \&
Buonanno 2007; Lousto \& Zlochower and references therein).  These
velocities far exceed the escape speed even from the brightest
galaxies and for this reason MBHs receding from their parent host have
been considered as electromagnetic targets of a coalescence event
(Devecchi et al. 2008; Loeb 2007; Volonteri \& Madau 2008).

Recently, the quasar SDSS J092712.65+294344.0 (hereafter SDSSJ0927;
Adelman--McCarthy et al. 2007) has been considered as first candidate
of a recoiling MBH by Komossa, Zhou \& Lu (2008, KZL).  SDSSJ0927
exhibits, in its optical spectrum, two distinct sets of lines offset
by 2650 km s$^{-1}$ relative to each other. The first set of narrow
emission lines (NELs) has redshift $z=0.713$ (identified by KZL as the
redshift of the galaxy that hosted the MBH at its birth). We will
refer to this set that is typical of an AGN as rest-frame NELs
(rf--NELs hereon; the red NELS in KZL).  The rf--NELs do not show any
evidence of ionization stratification (as in few other quasars),
possibly because these lines are not resolved in the SDSS spectrum.
The second set of lines comprises two blue--shifted systems featuring
different FWHM: the broad Mg II and Balmer emission lines (b--BELs
hereon) with FWHM $\approx 4000$ km s$^{-1}$, and the high-ionization
narrow lines (b--NELs) with FWHM $\approx 460-2000$ km s$^{-1}$, both
consistent with a redshift $z=0.698$. The b--BELs and b--NELs refer to
a line system in coherent motion relative to the rest-frame of the
host, with a light-of-sight velocity of 2650 km s$^{-1}$.  In KZL, the
permitted b--BELs result from a broad line emission system
gravitationally bound to the recoiling MBH, while b--NELs remains
problematic for the reasons explained in Section~\ref{sec:prob}.

The aim of this Letter is to suggest an alternative interpretation to
SDSSJ0927: that SDSSJ0927 is an unequal--mass MBHB embedded in a
circumbinary gaseous disc at the centre of its host galaxy.  In
Section ~\ref{sec:model} and ~\ref{sec:orb} we describe the main
properties of the model that are requested to reproduce the observed
properties of b--BELs.  In Section ~\ref{sec:gap} we discuss the
possible origin of b--NELs and of their Doppler shift.
In Section~\ref{sec:concl} we present our
conclusions.

\section{Limitations of the ejection hypothesis}\label{sec:prob} 

The origin of the b--NELs and the likelihood of observing a recoiling
MBH with a speed close to its maximum are the main concern of this
Section.

First, we notice that the recoil hypothesis by KZL seems unable to
explain all properties of the b--NELs observed in SDSSJ0927.  The FWHM
of the b--NELs is larger than the FWHM of the rf--NELs and of typical
FWHMs of NELs observed in AGNs.  By contrast, the FWHM of b--NELs is
not large enough to assure that the emitting gas is bound to the
recoiling MBH, although this should be the case as the b-NELs have the
same Doppler shift as the b-BELs.  KZL suggest that the b--NELs are
emitted by gas shocked in the interaction of the MBH with the
interstellar medium.  In this case however, there is no reason why the
speed of the shocked gas should equal that of the MBH (Devecchi et
al. 2008).  Alternatively, KZL consider that b--NELs are emitted in
the outer part of the accretion disc with substantial radial spreading
and/or in an unspecified kind of outflow.  This last possibility was
discussed in details by Bogdanovic et al. (2008).  They find that mass
outflows strongly reduce the density of the inner regions of the
accretion disc implying a low rate of gas accretion onto the MBH. In
this case the luminosity would be much weaker than observed, unless
the kicked MBH is more massive than $5 \times 10^9\, {\rm M}_{\odot}$.

\begin{figure}
\centering
\includegraphics [height=7 cm] {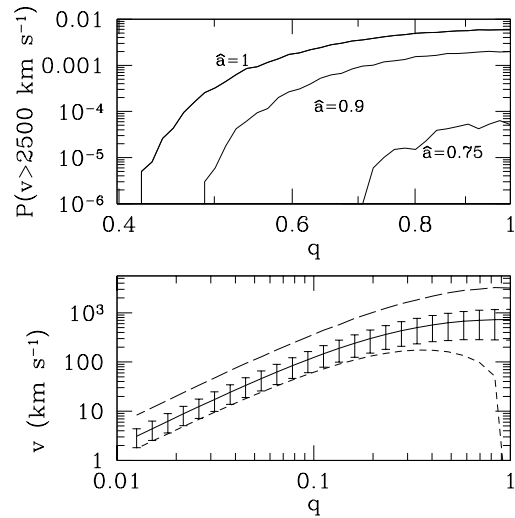}
\caption{\small{{\it Bottom panel}: recoil velocity as a function of
the binary mass ratio $q\equiv M_2/M_1$, for maximally Kerr holes
with $\hat {a}=1$. Long-dashed curve:
maximum recoil.  Short-dashed curve: minimum recoil. Solid curve: mean
recoil, with 1--$\sigma$ scatter. In all cases the recoil velocity is
averaged over $10^6$} orbital configurations for each $q$. {\it Top
panel:} probability for recoils faster than $v>2500\,{\rm km\,s^{-1}}$
as a function of $q$, for three representative spin
values (both holes are assumed to have the same spin), assuming
isotropic orbital parameters. No recoil with $v>2500\,{\rm
km\,s^{-1}}$ is possible for MBH spin
parameters ${\hat {a}}<0.65$.}
\label{fig:marta}
\end{figure}

Second, we remark that the probability of observing an ejected MBH
with the properties of SDSSJ0927 depends sensitively on the specific
frequency associated to MBHB mergers able to form a remnant with the
required large mass ($M_{\rm BH}$), and on the probability for the
remnant to receive such a high recoil velocity, i.e. $v>2500$ km
s$^{-1}$ (we conservatively adopt this lower limit to the recoil
velocity, which provides an upper limit to the detection
probability). The latter constraint strongly depends on the MBH spin
magnitudes and orientations with respect to the orbital plane, and on
the mass ratio of the binary, $q\equiv M_2/M_1\leqslant 1$ as
illustrated in Fig. 1. We model the recoil using the two available
fits to the recoil velocity based on the latest numerical relativity
results (Lousto \& Zlochower 2009).  Non-spinning holes always recoil
with velocities below 200$ \,{\rm km\,s^{-1}}$, and the recoil
increases with increasing spin.
In general the recoil is maximized for spin vectors lying in the orbital plane.

We can estimate the probability of detection of an ejection with
$v>2500\,{\rm km\,s^{-1}}$ by convolving the probability of large
recoils (shown in Fig.~1) with the merger rate of MBHBs.  We can
compute the number of MBH coalescences observable across the entire
sky, based on the merger rates by Volonteri, Haardt \& Madau (2003;
Volonteri \& Madau 2008). For a statistical sample of MBH mergers, our
models provide the masses of the merging MBHs and the redshift of the
event.  We select coalescences in the redshift range $0.5<z<1$
(compatible with the redshift of SDSSJ0927), further imposing a total
binary mass $M_{\rm BH}>5\times10^8\, {\rm M_\odot}$ (i.e. the mass of
the hypothetical recoiling MBH as discussed in KZL2008).
Marginalising over MBH masses and q, the highest rate of successful
recoils, $v>2500\,{\rm km\,s^{-1}}$, is found by imposing maximal
values for the dimensionless spin parameters, $\hat {a}=1$ for both
holes; the spins initially lie in the orbital plane pointing in
opposite directions.  The rate is $5\times 10^{-6}$ yr$^{-1}$.  The
requirement that the b-NELs come from a spreading accretion disc
however implies $M_{\rm BH}>5\times10^9\, {\rm M_\odot}$. For these
masses, the rate of events with $v>2500\,{\rm km\,s^{-1}}$ decreases
by one order of magnitude.  This set of assumptions, maximizing the
recoil over the orbital parameters, is mostly dependent on the mass
ratio $q.$ We need mass ratios $q\sim 1$ to obtain recoils above
$10^3\,{\rm km\,s^{-1}}$. Equal mass mergers are however rare (e.g.,
Volonteri \& Madau 2008; Sesana et al. 2005); the distribution of mass
ratios for $M_{\rm BH}\,\geqslant 5\times 10^8$ M$_{\odot}$ at
$0.5<z<1$ peaks at $q=0.3$ and decreases steeply at larger $q$.
Assuming MBHs with equal spins, $\hat{a}=1$, isotropically
distributed, and $M_{\rm BH}>5\times10^8\, {\rm M_\odot}$, the
expected rate decreases to less than $10^{-6}$ yr$^{-1}$.  The rate of
ejections decreases drastically with decreasing spin magnitude, and it
is zero for any orbital--spin configuration if $\hat {a}<0.65$.
\begin{figure}
\centering
\includegraphics [height=6 cm] {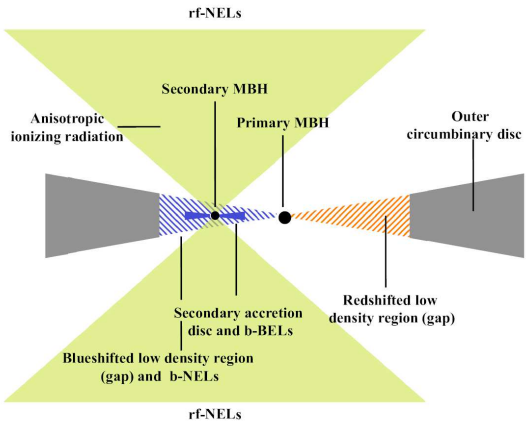}
\caption{\small{Sketch of the MBHB circumbinary disc structure in the
nucleus of SDSSJ0927.}}
\label{fig:sketch}
\end{figure}

\section{The MBH binary model}\label{sec:model}
In our model, SDSSJ0927 hosts a MBHB surrounded by a circumbinary thin
disc feeding the secondary MBH.  In the tidal interaction of the
binary with the gas a gap, i.e., a low density region surrounding the binary, opens.

It is known that, for MBHBs with $q\sim 1$,
$M_2$ is massive enough to perturb strongly the surrounding disc,
creating over few orbital periods an annular low density region,
commonly referred as ``gap''. During this phase most of the inner disc
is accreted by the primary (Syer \& Clarke 1995; Ivanov et al. 1999,
Dotti et al. 2006b), preventing long lasting accretion events onto
$M_1$.

As shown in Section 4, in order to explain the spectrum of SDSSJ0927,
$q \approx 0.3.$ Given this constraint, we expect that the tidal
interaction of the binary with the viscous disc 
will 
leave $M_1$ in a state of low activity.  
$M_2$, remaining
in closer gravitational contact with the inner rim of the outer disc,
moves at low speed relative to the higher density reservoir of gas,
thus sweeping up an accretion stream that inflows from the edge of the
gap. This continuous refueling supplies an accretion disc around
$M_2$, and allows for luminous accretion events during the orbital
decay of $M_2$ (Hayasaki, Mineshige, Sudou 2007; Cuadra et al. 2008).
The position of the inner edge of the outer circumbinary disc for $q \approx 0.3.$ 
(of interest for the scope of the paper) is known to lie between $\simeq
1.5-2 a $, where $a$ is the semi--major axis of the MBHB.  As for
large $q$ migration occurs on timescales much longer than the viscous
time in the outer disc (Ivanov et al. 1999), this binary can be
long-lived and we work under this hypothesis. A sketch of the model is
shown in Fig.~\ref{fig:sketch}.

We will consider $M_2$ the only accreting MBH of the binary.  In this
model, the rf--NELs are emitted from the ``standard'' large-scale
narrow line region that extends around the binary for $\sim 1$ kpc,
while the b--BELs are produced in the broad line region
gravitationally bound to $M_2$, and can be blue--shifted or
red--shifted depending on the orbital phase of the secondary. The
b--NELs come from a region of the gap near $M_2$, where forbidden
lines are emitted because of the low density of the gas. Since $M_2$
and the gas orbiting in the gap at comparable annuli are subject to
the same gravitational potential of $M_1$, both move with
approximately the same Keplerian velocity.  Accordingly, the same
blue--shift for the b--BELs and b--NELs can be explained if $M_2$
emits an anisotropic ionizing radiation normal to the plane of the
discs so that the ionizing photons interact only with the gas in the
gap near $M_2$. This is naturally achieved if, e.g., the accretion
flow around $M_2$ is a standard Shakura-Sunyaev accretion disc.
Indeed, cosine--like behaviour of the
thermal ionizing flux, coupled to the small solid angle subtended by
the disc assures that the ionization parameter of the gap gas not in
the immediate vicinity of $M_2$ is relatively low.

\section{Orbital parameters}\label{sec:orb}

In this section we use some of the observed properties of SDSSJ0927
(the blue--shift and the FWHM of the b--BEL, and the monochromatic
luminosity at 5100 \AA{}) to constrain the dynamical properties of the
hypothetical MBHB.  The compatibility of the model with the other
observed spectral features (in particular the blue--shift and the flux
of the b--NELs) is discussed in Section~\ref{sec:gap}. Given the large
blue--shift of the observed b--NELs and b--BELs, corresponding to a
high velocity of the accreting MBH, we assume that the active MBH is
$M_2$, the lightest hole, and that its mass is $6 \times 10^8$
M$_{\odot}$ as inferred in KZL2008. We will further assume that the
binary is moving on a (quasi) circular orbit.

The first constraint to be verified is that the observed broad line
region responsible for the b--BELs is bound to $M_2$, and is not
tidally stripped by $M_1$. This condition can be re--written as
$R_{\rm BLR}<R_2$, where $R_{\rm BLR}\approx 0.1$ pc (KZL2008) is the
radius of the BLR and $R_2$ is the Roche lobe radius of $M_2$. We estimate
$R_2$ using the approximation in Eggleton (1983), and impose:
\begin{equation} \label {eqn:eggleton}
\frac{R_{\rm BLR}}{a}<\frac{R_2}{a}= \frac{0.49 q^{2/3}}{0.6 q^{2/3}+ \ln(1+q^{1/3})}  
\end{equation}
where $a$ is the semi-major axis of the MBHB.

The observed velocity, $v_{\rm obs}$, of $M_2$ depends on $M_1$, $a$,
and the orientation of the plane of the binary in the sky:
\begin{equation} \label{eqn:keplero}
v_{\rm obs} = v_2 \sin i \cos \phi = \sqrt{\frac{G M_2}{a q (1+q)}} \sin i \cos \phi 
\end{equation}
where $v_2$ is the orbital velocity of $M_2$ relative to the centre of
mass, $i \in [0^{\circ},90^{\circ}] $ is the angle between the line of
sight and the direction normal to the orbital plane, and $\phi \in
[0^{\circ},360^{\circ}]$ is the phase of the orbit. From
eqs.~\ref{eqn:eggleton} and \ref{eqn:keplero}, and assuming $\cos \phi
=1$ in order to minimize $M_1$, we obtain
\begin{equation}\label{eqn:totale}
\frac{G M_2}{v_{\rm obs}^2 R_{\rm BLR}} \sin^2 i  >  q^{1/3}(1+q) \frac{0.6 q^{2/3}+ \ln(1+q^{1/3})}{0.49}. 
\end{equation}

The fraction on the left hand side of eq.~\ref{eqn:totale} is
constrained by the observations while the right hand side is a monotonically 
increasing function of $q$.  Given the inclination $i$,
eq.~\ref{eqn:totale} defines a maximum $q$, and a corresponding
minimum $M_1$. We focus on the minimum mass of $M_1,$ obtained for the
largest value of $\sin i,$ that is fitting all the observational
constraints.  Assuming that the unification model for AGNs (Urry \&
Padovani 1995) remains valid for this peculiar object, the observation
of the b--BELs limits the maximum value of the inclination of the
orbital plane to $i_{\rm max} \approx 40^{\circ}$, which implies
$q\approx 0.35$, $a\approx 0.34$ pc, and $M_1\approx 1.7 \times 10^9
{\rm M_{\odot}}$, value consistent with the observed high mass end of
the MBH mass function.
These values correspond to an orbital period of $370$ yr. The
time--scale for two MBHs in SDSSJ0927 to reach final coalescence due
to gravitational wave emission is $3 \times 10^9$ yr, shorter than the
Hubble time. If the binary model will be confirmed, this is proof of
the existence of a dynamical process able to overcome the so called
``last parsec problem'' and drive MBHBs to the final coalescence.

Using the FWHM of [OIII] as a tracer of the stellar velocity
dispersion, $\sigma_*$, and assuming the $M_{\rm BH}-\sigma_*$ relation 
(Ferrarese \& Ford, 2005 and references therein), we obtain for $M_1$ a mass $\sim 3$
orders of magnitude smaller than our estimate. This would be in contrast with the 
high luminosity of SDSSJ0927, which would imply extremely super Eddington 
luminosity ($\sim 10^3$ Eddington). On the other hand, in the MBHB scenario,
the dynamics of gas may be not representative of the galactic potential, 
due to the recent galaxy merger. 

To estimate the likelihood of the detection of a MBHB with the
properties inferred by our model, we use the same theoretical
merger-tree for cosmological MBHB used in Section~\ref{sec:prob}.  We
again impose limits on the redshift range ($0.5<z<1$), but now the
binary model constrains the masses of the binary members, rather than
the mass of the remnant. We therefore compute the probability that an
observer at $z=0$ detects binaries with $M_1>1.5\times 10^9 $
M$_{\odot}$ and $M_2>5\times 10^8 $ M$_{\odot}$ across the entire
sky. The observable merger rate corresponds to $\sim 10^{-4}$ binaries
per year, a value 20 times larger than the expected rate of ejections
obtained using the most favourable conditions for high kick velocity,
as discussed in Section~\ref{sec:prob}.

\section{Narrow emission from the gap}\label{sec:gap}  

Our model suggests that the b--NELs are produced from the low--density
gas in the gap region near $M_2$.  This assumption explains the
blue--shift of these lines, as we discussed in
Section~\ref{sec:model}.  For consistency, we need to constrain the
properties of the gas in the gap, and check if its physical status is
consistent with the production of b--NELs and with the energy flux
observed at these frequencies.

We assume that the gas in the gap is mostly ionized, so that we can
derive the gas properties from the luminosity of the Balmer lines.  Under the
hypothesis of isotropic emission the total luminosity of the H$\beta$
line is
\begin{equation}\label{eqn:hbeta}
L_{\rm H\beta}=4 \pi D_L^2 F_{{\rm H\beta}}=h \nu_{{\rm H\beta}} n_p n_e \alpha_{\rm H\beta} V_{\rm gap},
\end{equation}
where $D_L=4127$ Mpc is the luminosity distance of SDSSJ0927, $F_{{\rm
H}{\beta}}=1.9 \times 10^{-16}$ erg s$^{-1}$ cm$^{-2}$ is the flux of
the b--NEL component of the \Hb{} line (that we obtained directly by
fitting the rest-frame, de-reddened SDSS spectrum), $h$ is the
Planck constant, $\nu_{{\rm H\beta}}$ is the frequency of the \Hb{}
photons, $n_p$ and $n_e$ are the densities of protons and electrons respectively,
and $\alpha_{{\rm H\beta}}$ is the recombination coefficient.  In
order to estimate $n_e$ we need to constrain the volume of the
emitting region, $V_{\rm gap}$, that we assume to be half of the total
volume of the gap in which $M_2$ resides (see discussion in
Section~\ref{sec:model})
\begin{equation}\label{eqn:volume}
V_{\rm gap}=2 \pi \int_{0}^{R_{\rm gap}} r^2 \frac{H(r)}{r} dr,
\end{equation}   
where $H(r)$ is the thickness of the low density gaseous structure at
a radius $r$ and $R_{\rm gap} \approx 2 a$.  We use the standard
assumption that $H(r)/r=c_{\rm s}/v_{\rm Kepl}$, where $c_{\rm s}$ is
the local sound speed of the plasma and $v_{\rm Kepl}$ is the
Keplerian velocity of the gas due to the potential well of $M_1$.  In
this first estimate we assume $T\approx 10^4$ K, that corresponds to
$H(r)/r \approx 5 \times 10^{-3}$. Replacing this ratio in
eq.~\ref{eqn:volume} and using eq.~\ref{eqn:hbeta} we obtain $n_e
\approx 8 \times 10^6$ cm$^{-3}$.
 
A different estimate for $n_e$ can be obtained assuming that the
vertical support of the low density gas in the gap is due to a
supersonic turbulent motion of the plasma of the order of $1/2$ the
average FWHM of the b--NELs. In this second estimate we assume $v_{\rm
turb}\approx 200$ km s$^{-1}$, corresponding to $n_e \approx 2 \times
10^6$ cm$^{-3}$.  We emphasize that these estimates must be
considered as a lower limit, because the gas can be clumpy (so that
only a fraction of the irradiated gas will contribute consistently to
the \Hb{} emission), and because the anisotropic radiation produced in
the  accretion disc of $M_2$ can irradiate less than 1/2 of the gap.
 
Both estimates of $n_e$ are much smaller than the density expected for
a standard $\alpha$--disc around $M_1$ at $r \sim a$, indicating the
presence of a gap, confirming our expectations.  An additional,
independent constraint on $n_e$ can be obtained from the flux ratios
of different Oxygen lines, in particular the [OII]$_{3727}$ doublet
and the [OIII]$_{4363, 4959, 5007}$ lines (see e.g. Osterbrock \&
Ferland 2006 and references therein).  The ratio between the two
components of the [OII] doublet cannot be employed here, since the
lines are practically unresolved. We restrict our analysis to the
[OIII] lines. Since the b-[OIII]$_{5007}$ and the rf-[OIII]$_{4959}$
are blended, we assume a factor of 3 between the fluxes of the
[OIII]$_{5007}$ and [OIII]$_{4959}$ lines in both systems (Osterbrock
\& Ferland 2006). We compare the observed line flux ratios to the
expectations for a gas in photo-ionization equilibrium, which yields
$T_e=10^4-2\times 10^4$ K.  For $T_e=10^4$ K, we obtain $n_e =
(2.5_{-0.5}^{+1.0}) \times 10^6$ cm$^{-3}$. This density is fully
consistent with the predictions from the H$\beta$ luminosity.
Increasing $T_e$, $n_e$ decreases.  At the maximum temperature allowed
by the photo-ionization equilibrium, $2\times 10^4$ K, the b-system
has $n_e=(4_{-1}^{+2}) \times 10^5$ cm$^{-3}$.  Typically [OIII] lines
are produced in regions with much lower density (narrow line regions,
$n_e\simeq 10^3$ cm$^{-3}$).  SDSSJ0927 is indeed confirmed to be a
peculiar system.

We check our results by calculating if the production of a forbidden
line such as [OIII] in such high density regions is self-consistent.
We therefore consider the contribution of both collisional excitations
and de--excitations when computing the luminosity of [OIII]
lines. This calculation depends on the volume of the emitting region,
on $n_e$, and on the density of [OIII] ions, $n_{\rm [OIII]}$.  Via
detailed equilibrium balance, with the temperatures and electron
densities derived above, we find $n_{\rm [OIII]} \simeq$ 2--4 orders
of magnitude lower than $n_e$.  This ratio is fully consistent with
standard metallicities in quasars hosts.

This consideration can explain the presence of both H$\beta$ and
forbidden lines, but the [OII] doublet.  [OII] emission is not
consistent with such high densities, because the [OII] critical
density for collisional de-excitation is $n_e \lsim 10^4$ cm$^{-3}$.
The detection of [OII] lines suggests that the gap is filled with an
inhomogeneous, cloudy medium. The source of this inhomogeneity may be
the competition between the gas inflow from the circumbinary disc and
the periodic perturbation exerted by the MBHB (see, e.g. Hayasaki,
Mineshige, Sudou 2007; Cuadra et al. 2008; Bogdanovic et al. 2008).
In an inhomogeneously filled gap, \Hb ~lines are emitted in the
highest density regions ($n_e \gsim 2\times 10^6$ cm$^{-3}$), while
[OIII] and [OII] lines are produced by intermediate ($3 \times 10^5$
cm$^{-3}$ $ \lsim n_e \lsim 3.5 \times 10^6$ cm$^{-3}$) and low ($n_e
\lsim 10^4$ cm$^{-3}$) density regions, respectively.

\section{Discussion and conclusions}\label{sec:concl}

The binary hypothesis for SDSSJ0927 relies on a simple set of
assumptions.  We can relax the assumption of circular orbits, and
assume a non zero eccentricity ($e<0.5$), expected to be present after
gap opening (Armitage \& Natarajan 2005; Dotti et al. 2006a, 2007;
Cuadra et al. 2008).  Eccentricities of this magnitude are still
compatible with the binary model.  Larger values are instead unlikely
in SDSSJ0927: if the binary has a large eccentricity and $M_2$ is near
apocentre, an extremely massive $M_1$ is needed to justify such a
large orbital velocity.  On the other hand, Shields et al. (2009) set
an upper limit on the change of the Doppler shift of the b-BEL system
($<10$ km/s/year). In the binary scenario, this prevents $M_2$ to be
caught too close to the pericentre. We note that assuming circular
orbit and the set of MBHB parameters discussed above, the b-BEL
system moves only of 4.2 km/s over three years, corresponding to an
average shift of $\approx 1.4$ km/s/year.

SDSSJ0927 is clearly exceptional. Either it is the first example of a
MBH ejected from its nucleus due to the gravitational recoil, or it
harbours one of the smallest separation MBHBs ever detected. Only the
possible binary in OJ287 could have closer MBHs ($\sim 4.5\times
10^{-2}$ pc, Valtonen 2007). SDSSJ0927 would provide clear evidence
that MBHBs can indeed reach coalescence in less than a Hubble time,
and that Nature is able to solve the last-parsec problem. If the model
discussed in this letter is confirmed, SDSSJ0927 will represent the
first detection ever of emission from a circumbinary gap around a
MBHB.

The proof of existence of binaries with such a small orbital
separation is of extreme importance for gravitational wave experiments
such as LISA (Bender et al. 1994), and the Pulsar Timing Arrays (PTA,
Sesana et al. 2008).  Since our theoretical expectations for a binary
with the characteristics of SDSSJ0927 are of order $10^{-4}$ (binary
hypothesis) down to $10^{-6}$ (ejection hypothesis) per year, either
we have been extraordinarily lucky, or MBHBs are {\it more common}
than predicted, opening exciting possibilities for LISA and PTA.

\section*{Acknowledgments}
The authors thank Michael Eracleous for making them aware of a similar
paper that independently reaches similar conclusions on SDSSJ0927.  We
are also grateful to Bernadetta Devecchi, David Hummer, Ruben
Salvaterra, and Pete Storey for fruitful discussions. Support for this
work was provided by grant SAO-G07-8138 C (M.V.).

\label{lastpage}

\end{document}